\documentclass[pre,aps,10pt,twocolumn,superscriptaddress,notitlepage,graph]{revtex4-1}
\usepackage{bbm,bm}
\usepackage{graphicx}

\newcommand{\beq}{\begin{equation}}
\newcommand{\eeq}{\end{equation}}

\def\DAB{\Delta_{\rm AB}}

\newcommand {\tw}	{t_\mathrm{w}}

\newcommand {\TG}	{T^{*}}

\newcommand{\caja}[1]{\left[ {#1} \right] }
\newcommand{\paren}[1]{\left( {#1} \right) }

\newcommand{\av}[1]{{\left\langle {#1} \right\rangle}}

\renewcommand{\vec}[1]{\boldsymbol{#1}}

\begin{document}
\title{Low-temperature anomalies of a vapor deposited glass}

\author{Beatriz Seoane}
\affiliation{Laboratoire de physique th\'eorique, D\'epartement de physique de
l'ENS, \'Ecole normale sup\'erieure, PSL
Research University, Sorbonne Universit\'es, CNRS,
75005 Paris, France}
\affiliation{Instituto de Biocomputaci\'on y F\'{\i}sica de Sistemas Complejos (BIFI), 50009 Zaragoza, Spain}

\author{Daniel R. Reid}
\affiliation{Institute for Molecular Engineering, University of Chicago, 5640 South  Ellis Avenue Chicago, Illinois 60637}

\author{Juan J. de Pablo}
\affiliation{Institute for Molecular Engineering, University of Chicago, 5640 South  Ellis Avenue Chicago, Illinois 60637}

\author{Francesco Zamponi}
\affiliation{Laboratoire de physique th\'eorique, D\'epartement de physique de
l'ENS, \'Ecole normale sup\'erieure, PSL
Research University, Sorbonne Universit\'es, CNRS,
75005 Paris, France}

\begin{abstract}
We investigate the low temperature properties of two-dimensional Lennard-Jones glass films, prepared {\it in silico} both by liquid cooling and by physical
vapor deposition. We identify deep in the solid phase a crossover temperature $\TG$, at which slow dynamics
and enhanced heterogeneity emerge. Around $\TG$, localized defects become visible, leading to vibrational anomalies as compared to 
standard solids. We find that on average, $\TG$ decreases in samples with lower inherent structure energy, suggesting that such anomalies will be suppressed in ultra-stable
glass films, prepared both by very slow liquid cooling and vapor deposition. 
\end{abstract}

\maketitle

Low-temperature crystalline solids are usually described in terms of
harmonic vibrations around a perfect periodic lattice (phonons).  Within this framework, defects such as vacancies and dislocations can be treated as small perturbations.
This description breaks down for amorphous solids such as glasses, foams,
emulsions, plastics, colloids, granular materials, bacterial colonies,
and tissues~\cite{FL98,SWS07,HKLP10,PRB16,KHGGC11,CWKDBHR10,CSRMRDKL15}.  In these systems, the identification
of ``defects'' becomes challenging because the solid ground state is strongly disordered. 
As a consequence,
amorphous solids display many universal {\em anomalies} with respect to crystals. Examples are the so-called Boson Peak, an excess
of low-energy vibrational modes~\cite{MS86}; the anomalous scaling of heat capacity and thermal conductivity with temperature~\cite{ZP71,Ph87}; the irreversible plastic
response to arbitrarily small perturbations~\cite{FL98,ML99,HKLP10,PRB16}; and highly cooperative
relaxation dynamics, contributing to the so-called $\beta$-processes~\cite{HB08,Go10,CPPWN12}. 

These anomalies have been widely reported in amorphous solids of very
different nature.  Interestingly, recent experimental work has shown
that by preparing glasses through a process of physical vapor
deposition, one can produce ultra-stable states that lie deep in the
free energy landscape~\cite{Sw07}.  Compared to their liquid-cooled
counterparts, vapor-deposited glasses show higher density
\cite{Dalal2012a} and kinetic stability \cite{Sw07,Leon2010}.  When
these ultra-stable glasses are studied at very low temperatures, it is
found that the anomalies characteristic of amorphous solids are
strongly
supressed~\cite{QLKMH13,perez2014suppression,LQMKH14,yu2015suppression,TCBSE16}.

Many theoretical approaches to this problem are based on the study of the potential energy landscape of 
glass-forming particle systems~\cite{Go69,SW83,sastry1998signatures,debenedetti2001supercooled,heuer2008exploring,Go10,CKPUZ14n}.
These studies have suggested that glass anomalies can be interpreted in terms of glass states being
not well-defined energy minima, but structured metabasins containing a collection of sub-basins separated by barriers
of variable size~\cite{RH04,MW01}, see Fig.~\ref{fig:sketch}. 
In particular, recent work~\cite{CJPRSZ15,berthier2015growing,camille2017} has identified a set of simple observables
(the mean square displacement between identical ``clones'' of the original system) that allows one to detect easily
the development of a structure of sub-basins inside a glass metabasin.

In this work, using the methods of~\cite{CJPRSZ15,berthier2015growing,camille2017}, 
we explore {\it in silico} the potential energy
landscape of binary Lennard-Jones glass films prepared through two
experimentally relevant protocols: slow liquid cooling, and physical
vapor deposition following Ref.~\cite{reid2016age}.  
We study these films due to their experimental relevance and the fact that 
they have been well characterized by previous work~\cite{reid2016age,reid2016b}.  
In contrast to
previous studies which prepared bulk equilibrium samples using the swap
algorithm~\cite{berthier2015growing,camille2017}, our
film preparation methods --inspired by the vapor-deposition experimental protocol-- produce
non-equilibrium films that are expected to be higher in the
potential energy landscape than experimentally prepared vapor-deposited 
glasses~\cite{NBC17,BCFZ17}. 
In addition, both our liquid-cooled and vapor-deposited films are
prepared in the presence of both a substrate and a free surface,
allowing the study of these features' influence on the low-temperature
physics of the samples.  

We find that a threshold $\TG$ can be detected within the glass phase, below which vibrational dynamics of the solid become
orders of magnitude slower and the structure of the glass basin becomes visible. The value of $\TG$ depends primarily on film stability,
decreasing substantially
with the inherent structure energy of the sample - a measure of stability~\cite{reid2016b} -, while a protocol dependence of $\TG$ is not
detected.
This observation
is compatible with the disappearance 
of anomalies in ultra-stable glasses.
Furthermore, we observe
significant sample-to-sample variations both in the value of $\TG$
and in the aging dynamics below this
threshold. 
All samples display localized defects, however several samples
display collective dynamics, which could be related to cooperative displacements enabled by the free surface.
It is important to note that the glasses considered here incorporate the non-equilibrium nature of real materials, as well as the presence of a substrate and free boundary, which have an important impact on the physics below $\TG$.  Note also that the films considered
in this study have fixed thickness (the same used in Ref.~\cite{reid2016age}), so the dependence of the results on films' thickness is not addressed here and left for future work.

\section{Sample preparation protocol}

\begin{figure}[t]
  \centering
  \includegraphics[trim=0 240 0 270,width=\columnwidth]{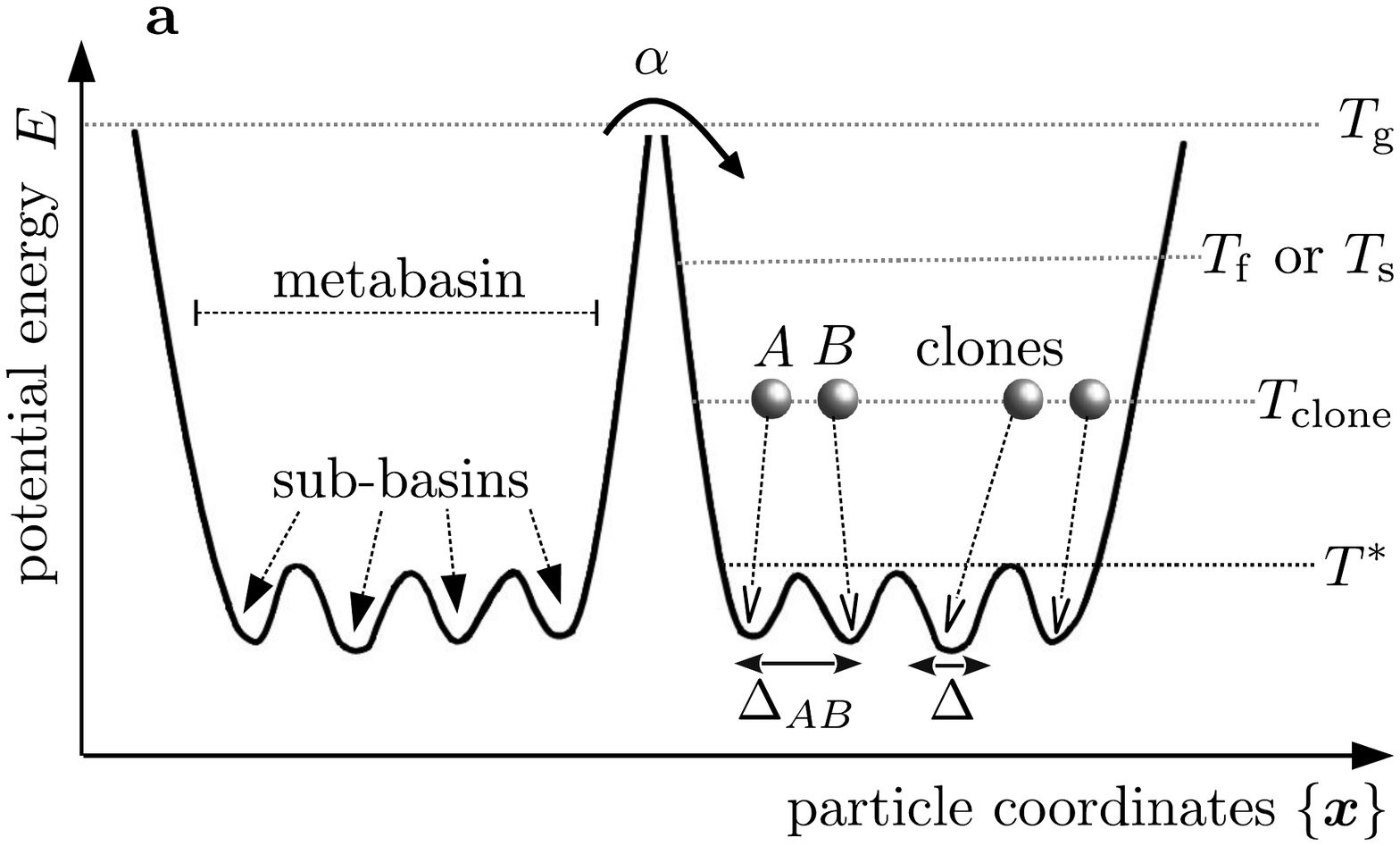}
  \caption{{\bf Illustration of the energy landscape.} 
  Samples are prepared in a glass basin at temperature $T_{\rm f}$ or $T_{\rm s}$ lower than the glass transition $T_{\rm g}$.
  They are then brought at lower temperature $T_{\rm clone}$ to ensure that no residual diffusion ($\alpha$-relaxation) is present. There, clones are produced to sample
  the interior of glass basins. Our main observables are $\Delta(t,\tw)$, the mean square displacement of particles in a single clone, and $\Delta_{\rm AB}(\tw)$, the 
  displacement between two distinct clones.
  \label{fig:sketch}}
\end{figure}

Here we provide a brief description of the system and protocols used
in this work.  See Ref.~\cite{reid2016age}, the illustration in
Fig.~\ref{fig:sketch} and the Appendix for details.  We prepare
$N_{\rm s}$ glass {\it samples} of a binary two-dimensional
Lennard-Jones system which shows a glass transition temperature close
to $T_{\rm g} \approx 0.21$ for the range of cooling rates used in
this study.  We study two distinct classes of films: {\it (i)}~those
formed by slow cooling~(SC) of liquid films into the glass phase at a
final temperature $T_{\rm f}$ with two distinct cooling rates
$\delta_{\mathrm{SC}}$, and {\it (ii)}~those formed by a procedure
mimicking physical vapor deposition~(VD).  In VD, we use four
different deposition rates $\delta_{\mathrm{VD}}$ with substrates
held at temperature $T_{\rm s}$.  In both protocols, $T_{\rm f} <
T_{\rm g}$ or $T_{\rm s} < T_{\rm g}$, so the samples we produce are
in the glass phase.

\begin{figure}[t]
  \centering
  \includegraphics[trim=0 10 0 0,width=\columnwidth]{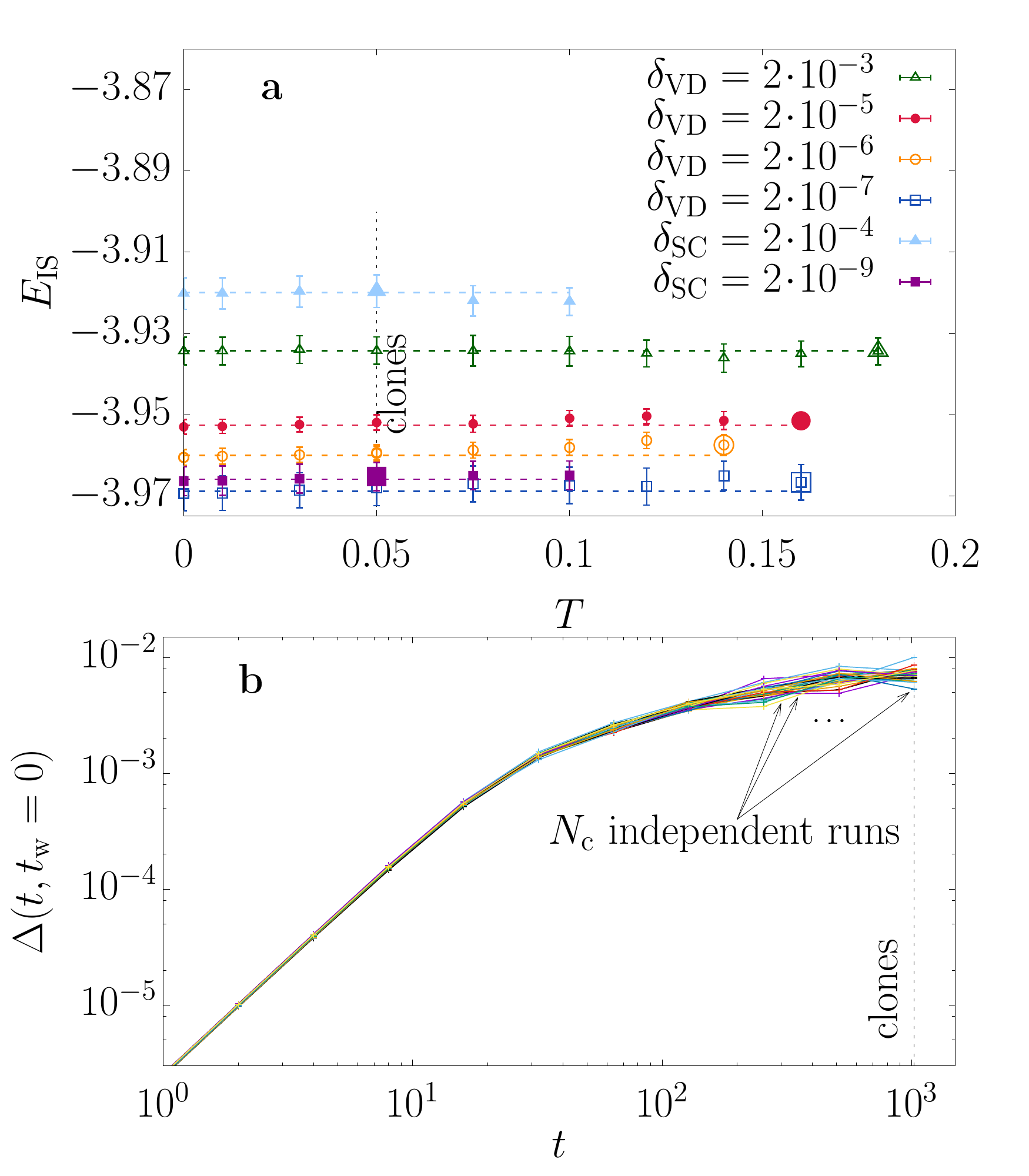}
  \caption{{\bf Preparation of glass samples and clones.} 
  (a)~Inherent structure energy
    $E_\mathrm{IS}$ measured at $T_{\rm f}$ or $T_{\rm s}$ (larger symbol), and upon cooling or heating the samples at different $T$. 
    Dashed lines are constant fits to the data for $T\leq T_{\rm clone}=0.05$, that demonstrate that the system is not leaving its metabasin.
    (b)~$\Delta(t,\tw=0)$ at temperature $T= T_{\rm clone}$ for
    one of the VD samples prepared with
    $\delta_\mathrm{VD}=2 \cdot 10^{-6}$. 
    The length of the clone
    preparation simulations is indicated by the dotted line, whose length is the
    size of the cage (i.e.~the height of the plateau).
  \label{fig:eIreid}}
\end{figure}

To study the vibrational anomalies of a glass basin, each sample is
brought to a lower temperature $T_{\rm clone}$ (lower than either
$T_{\rm f}$ or $T_{\rm s}$).  The same $T_{\rm clone}$ is employed in all cases,
i.e. for
all samples and all protocols.  At this temperature the system is sufficiently close to
its inherent structure. In fact, as shown in Fig.~\ref{fig:eIreid}a,
the inherent structure energy (as computed by energy minimization
configurations at different temperatures) remains constant below this
temperature for all the glasses considered.  
From this, we suggest that no diffusion occurs over this period.  
We verify that the system behaves as
a normal solid at $T_{\rm clone}$, meaning that the state is ergodic
and the vibrations of the particles are weakly correlated.  Once
cooled, we prepare $N_{\rm c}$ {\it clones}, or independent
configurations distributed within the basin of each glass sample. In
practice, each clone is obtained as the result of an independent simulation of
length $t_\mathrm{clones}$, the dotted line in Fig.~\ref{fig:eIreid}.
The clones are then instantaneously quenched to a final temperature
$T < T_{\rm clone}$ and their dynamics are examined at constant $T$,
with $\tw$ being the time elapsed since the quench. Note that when samples
are studied at  $T = T_{\rm clone}$, the dynamics are stationary, and for this reason
the origin of time can be chosen arbitrarily.

Following previous work~\cite{CJPRSZ15,berthier2015growing,camille2017},
we focus our attention on two observables: 
\beq \Delta (t+\tw,\tw) = \frac{1}{N} \sum
_{i=1}^{N} \overline{\av{|\vec{r}_i(t + \tw) - \vec{r}_i(\tw)|^2}} \ ,
\label{eq:MSD}
\eeq
which is the mean square
displacement of particles in each clone between time $\tw$ and $t+\tw$, and
\beq \Delta_{\rm AB} (\tw) = \frac{1}{N} \sum
_{i=1}^{N}\overline{ \av{|\vec{r}_i^A(\tw) - \vec{r}_i^B(\tw)|^2} }\ .
\label{eq:DAB}
\eeq
which is the mean square displacement between particles in
two distinct clones (denoted A and B) of the same sample at the same time  $\tw$,
  $\{\vec{r}_i^A(\tw)\}$ and $\{\vec{r}_i^B(\tw)\}$.
Here, $\av{\ \bullet \ }$ refers to the
thermal average, computed as the average over all
the clones of the same sample, while $\overline{\ \bullet \ }$ refers to 
the average over all the samples with
the same preparation procedure. To increase the
statistics, the thermal average of $\Delta_{\rm AB}$ is computed using all
the $N_\mathrm{c}(N_\mathrm{c}-1)/2$ possible couples of A and B
clones, but the error bars are computed by taking into account the correlations
between pairs using the jack-knife method~\cite{amit2005field}.

Both quantities are computed for particles in the 
middle region of the sample (the region in between the two horizontal lines
in Fig.~\ref{fig:snapshots}). In this region the density and relative concentration of the two particle types are both constant \cite{reid2016age}, allowing boundary effects to be avoided. The displacement of the center
of mass of the whole sample is removed. Both observables are averaged over clones and, unless
otherwise specified, over multiple samples prepared with the same protocol.

\section{Results}

\subsection{Clones are prepared in an ergodic state} 

\begin{figure}[t]\centering
  \includegraphics[width=\columnwidth]{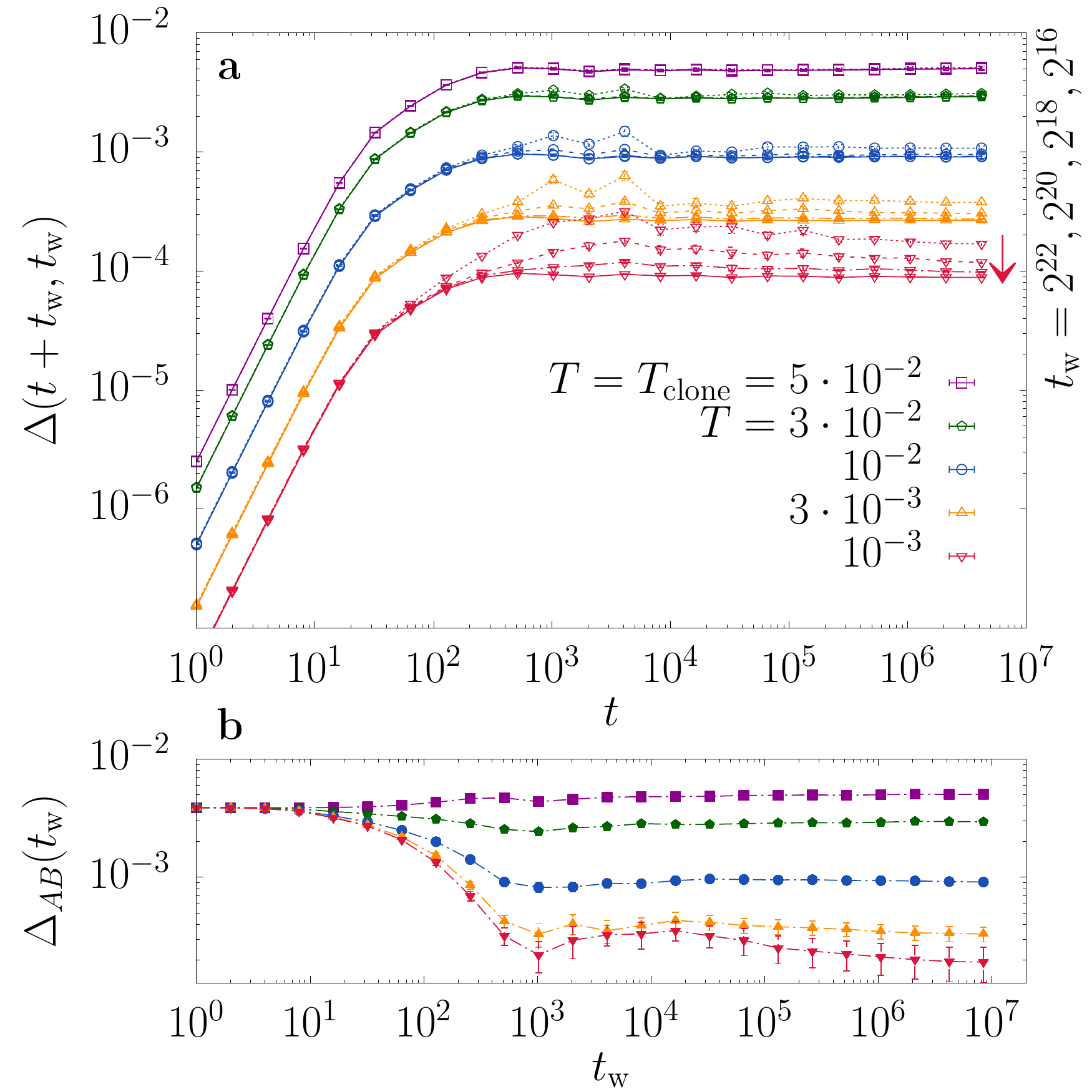}
  \caption{{\bf Emergence of slow vibration dynamics.} (a)~Aging
    effects in $ {\Delta(t+\tw,\tw)}$ for VD glasses obtained with
    $\delta_\mathrm{VD}=2\cdot 10^{-6}$. For each temperature we
    plot four values of $\tw$ ($2^{16},2^{18},2^{20},2^{22}$), all of them corresponding to the regime where
the ballistic part has converged to the same curve. (b) Aging effects in 
    $\Delta_\mathrm{AB}(\tw)$.  
  }\label{fig:aging}
\end{figure}

\begin{figure}[t]\centering
  \includegraphics[width=\columnwidth]{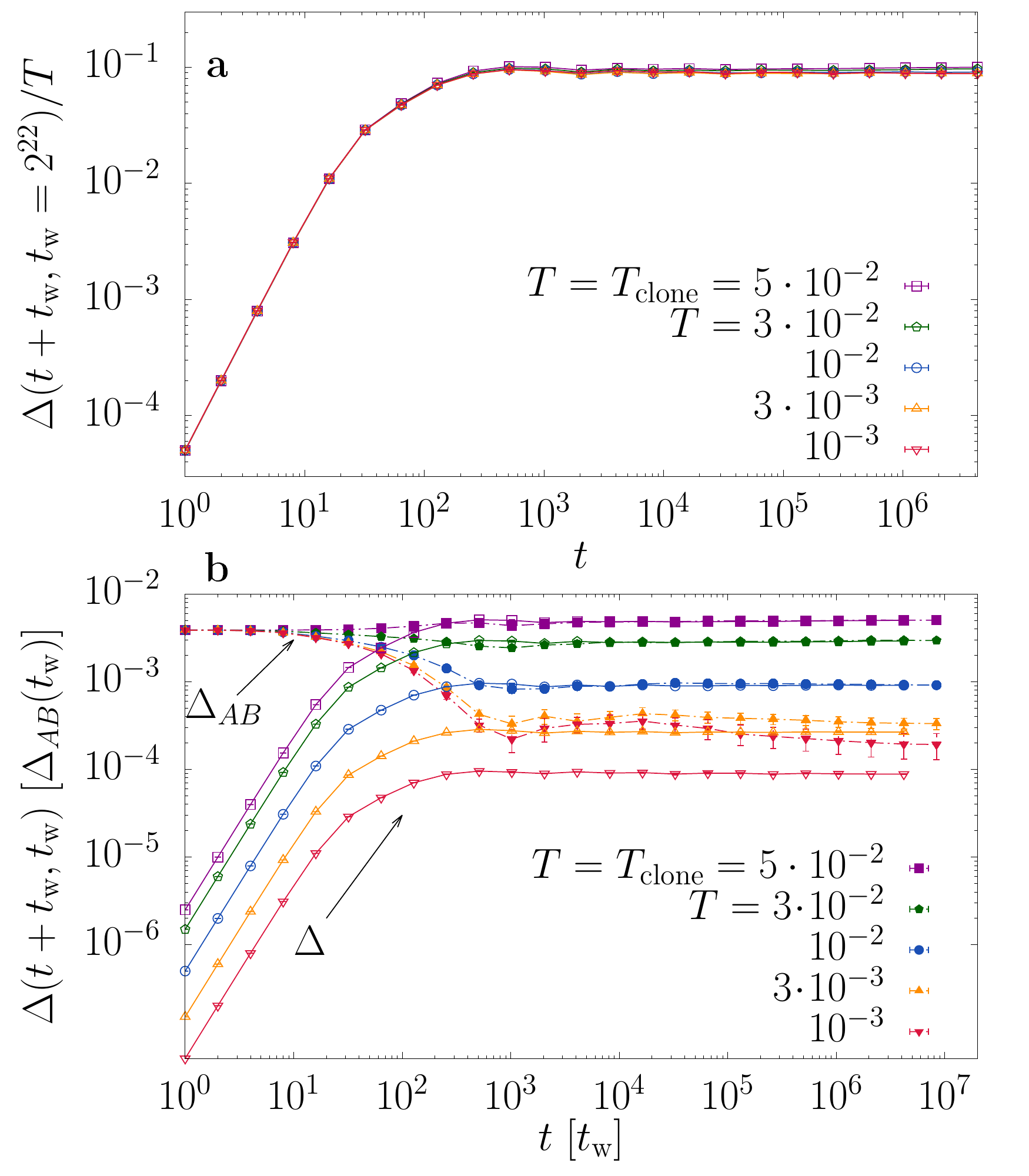}
  \caption{{\bf Breakdown of the ergodicity.} 
    (a) The scaled $\Delta(t+\tw,\tw)/T$ for the largest $\tw = 2^{22}$, at different temperatures. There is an excellent
    collapse indicating that for large $\tw$ the MSD is proportional to temperature.
    (b)~We compare $\Delta(t+\tw,\tw)$ for $\tw=2^{22}$ vs. $t$ with
    $\Delta_\mathrm{AB}(\tw)$ vs. $\tw$ of Fig.~\ref{fig:aging}. The onset where
    $\Delta_{\rm AB}^\infty = \Delta_{\rm AB} (\tw \to \infty) \neq\Delta (t \to \infty, \tw
    \to \infty) = \Delta^\infty$ corresponds to the emergence of aging of Fig.~\ref{fig:aging}a.}\label{fig:aging2}
\end{figure}

We begin by
discussing the behavior of $\Delta (t+\tw,\tw)$ and
$\Delta_{\rm AB} (\tw)$ for samples at the clone preparation
temperature $T_{\rm clone}$ (see purple squares in Fig.~\ref{fig:aging}).
Because clones have been prepared well below $T_{g}$, no diffusion is
observed in our simulation time windows, meaning that the averaged
cage size $\Delta^\infty$ of the material at each temperature can be
extracted from the plateau value of $\Delta(t+\tw,\tw)$ at long $\tw$,
as shown in Fig.~\ref{fig:aging}a. On the other hand, 
$\Delta_{\rm AB} (\tw)$ reach a constant value at long times, see Fig.~\ref{fig:aging}b, that precisely coincides with $\Delta^\infty$ (it can be better appreciated in  Fig.~\ref{fig:aging2}b where both observables are plotted superimposed).
The convergence of 
these two quantities in the long time limit means that a single trajectory
of the system samples, at long times, the same states that are sampled by two
independently prepared clones.  This indicates that the glass basin is comprised of well-defined internal cages which are ergodically sampled, and that vibrations of particles remain weaky correlated~\cite{berthier2015growing,seguin2016,camille2017}.

\subsection{Growing timescales upon cooling} 

\begin{figure}[t]
\centering
  \includegraphics[width=\columnwidth]{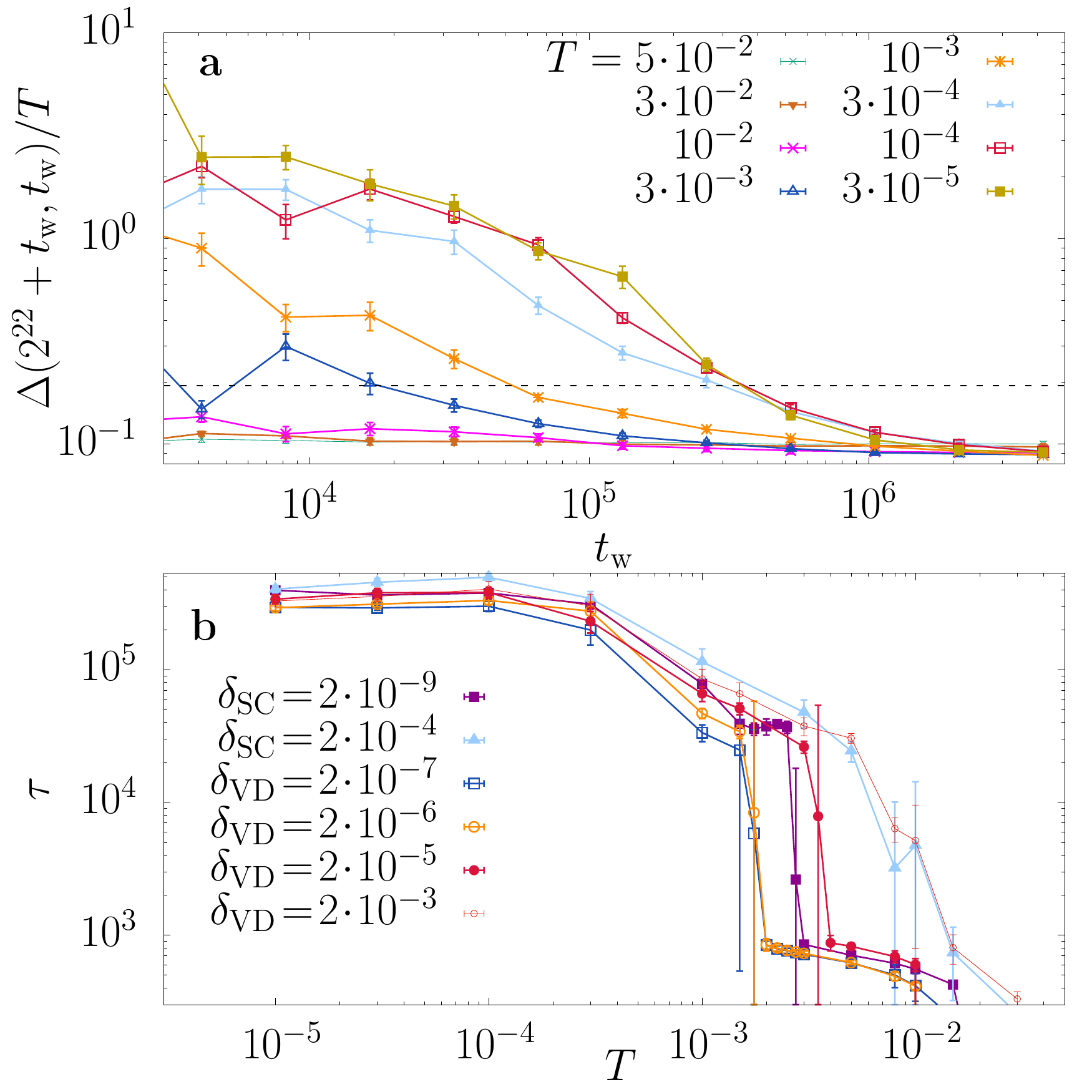}
  \caption{{\bf Convergence to the plateau.} (a) The plateau
    height (long $t$ limit of $\Delta (t+\tw,\tw)/T$, estimated by the largest available $t$) 
    as function of $\tw$ for
    different temperatures, extracted from the data of Fig.~\ref{fig:aging2}a.
    (b) The time $\tau$ needed to fall below the dashed line in
    (a), as a function of $T$ for differently prepared glasses.
  }\label{fig:agingtau}
\end{figure}

Next, we study the behavior
of $\Delta (t+\tw,\tw)$ and $\DAB(\tw)$ as a function of $t$, using different
reference times $\tw$ elapsed after a sudden drop in temperature from
$T_{\rm clone}$ to $T<T_{\rm clone}$ (Fig.~\ref{fig:aging}).  At small
values of $\tw$, one expects a sharp nonequilibrium response of
$\Delta (t+\tw,\tw)$ and $\DAB(\tw)$ to the change of temperature:
this is manifested both at small $t$, during the ballistic exploration
of the cages, and at long $t$ in the plateau region. The value of the
plateau evolves from the typical cage size at the preparation
temperature (at very small $\tw$), to the new temperature one (longest
$\tw$), see Fig.~\ref{fig:aging}. In addition, as we show in
Fig.~\ref{fig:aging2}a, the limiting long-$\tw$ curves at different
temperatures can be roughly collapsed in a single curve by dividing
them by the temperature, which suggests that the size of the cages
increases linearly with temperature.  However, the typical time it
takes to the system to converge to this long-$\tw$ plateau, depends
drastically on the final temperature. This is more clearly seen by
plotting $\Delta (t+\tw,\tw)/T$ for $t=2^{22}$ [i.e. the long $t$
limit of the mean-squared displacement (MSD)] as function of $\tw$, see
Fig.~\ref{fig:agingtau}a. While at high temperatures the plateau
converges rapidly to its final value, this convergence slows
significantly as the temperature is decreased. In order to quantify
this effect, we extract the time $\tau$ such that for $\tw > \tau$,
the value of $\Delta (2^{22}+\tw,\tw)/T$ is consistently below a
threshold (dashed line in Fig.~\ref{fig:agingtau}a). We fixed the
threshold to 0.2 for all the samples.  The errors are obtained using
the jack-knife method~\cite{amit2005field}. We show $\tau$ as a
function of $T$, for glasses prepared by different protocols in
Fig.~\ref{fig:agingtau}b, finding that these characteristic times grow
very quickly in the vicinity of well defined temperatures that depend
on how the material was prepared. Of course, within this approach, the values of
$\tau$ depend of the threshold chosen, and we observed that the
temperatures at which the sharp growth occurs also shift
mildly (effect included in the error bars). Nevertheless, the overall picture remains the same.

Furthermore, the convergence to the final
cage sizes (Fig.~\ref{fig:aging}a) slows at roughly the same
temperatures at which $\Delta (t+\tw,\tw)$ and $\Delta_{\rm AB} (\tw)$ 
no longer converge to the same plateau value at long times, as
shown in Fig.~\ref{fig:aging2}b. The large time limit of
both quantities, which we call $\Delta^\infty$ and $\Delta_{\rm AB}^\infty$, plotted as a function
of $T$ (Fig.~\ref{fig:deltaandTg}b), 
converge to the same values at high
temperatures ($T\gtrsim10^{-2}$), while they clearly separate at low temperatures. 
It is however important to note that the relaxation time $\tau$ (Fig.~\ref{fig:agingtau}b)
does not seem to diverge at any finite temperature: instead, it saturates. 
One may wonder whether this saturation is simply due to the finite size of the system, in which case 
the value of $\tau$ at low temperature would increase with system size. Ruling out this possibility would require
a careful finite size study, that we leave for future work.

\subsection{Temperature threshold and inherent structure energy} 

\begin{figure}[t]
\centering
    \includegraphics[width=\columnwidth]{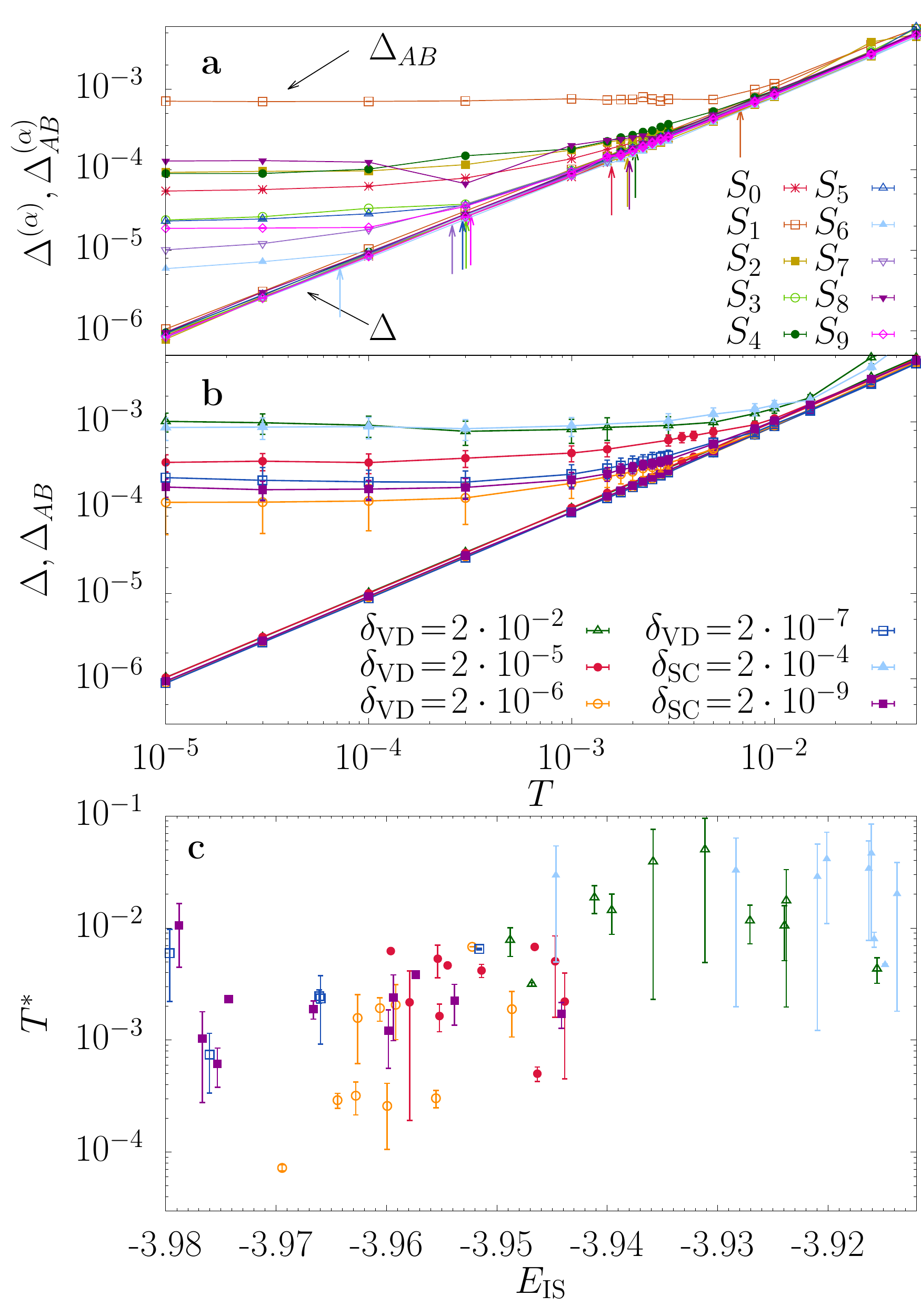}
  \caption{  {\bf Exponential decay of $\TG$ with the inherent structure energy of the samples. } 
  (a)~Individual long time values of $\Delta^\infty$ and $\Delta^\infty_{\rm AB}$, plotted versus $T$, for the 10 VD glass samples obtained with
    $\delta_\mathrm{VD}=2\cdot 10^{-6}$. The separation of $\Delta^\infty$ and $\Delta^\infty_{\rm AB}$ happens at a strongly
    sample-dependent temperature.
(b)~Averages of $\Delta^\infty$ and $\Delta^\infty_{\rm AB}$ over samples, plotted versus $T$, for different sample preparation protocols, showing that
the average separation temperature depends on the preparation protocol and decreases for slower protocols.
(c)~Values of $\TG$ as function of the inherent structure energy, as extracted sample by sample for the different preparation protocols (different colors with the same
color code as in panel~b), 
showing a high correlation between $E_\mathrm{IS}$ and $\TG$. 
  }\label{fig:deltaandTg}
\end{figure}

We have seen
that the long time limits $\Delta^\infty$ and $\Delta_{\rm AB}^\infty$ separate
near a threshold $\TG$, indicating a loss of ergodicity within
the glass basin below this temperature, which is also associated with
the emergence of much slower aging dynamics. If we examine each
sample individually, we find that the value of $\TG$ fluctuates
strongly from sample to sample (Fig.~\ref{fig:deltaandTg}a),
 and it depends strongly on how the sample was prepared, as we show in Fig.~\ref{fig:deltaandTg}b by taking the sample averages. 

 To study systematically the dependence of $\TG$ on sample
 preparation method and rate, we define it more precisely as follows.
 We compute the temperature below which $\Delta^\infty$ and
 $\Delta_\mathrm{AB}^\infty$ become distinct in each sample, and the
 temperature for which
 $\Delta^\infty(T)=\Delta^\infty_\mathrm{AB}(T= 10^{-5})$, i.e. the
 point at which a horizontal line equal to the zero-temperature value
 of $\Delta_{\rm AB}^\infty$ intersects $\Delta^\infty(T)$.  We define
 $\TG$ as the average of these two estimations (indicated by
 the arrows in Fig.~\ref{fig:deltaandTg}a), and we associate to it an
 error given by half the difference of these two estimations. The
 reason for this is that $\TG$ does not correspond to a sharp
 phase transition but rather to a crossover, therefore one cannot
 define $\TG$ unambiguously.  We show the results for the
 $\TG$ of each sample in Fig.~\ref{fig:deltaandTg}c as function
 of their inherent structure energy, which is correlated with the
 cooling or deposition rate and is a proxy for glass
 stability~\cite{reid2016age,reid2016b}.  In spite of the large spread
 of the data points, we find that the values of $\TG$ are
 correlated (the linear correlation coefficient is 0.67) with the
 logarithm of the inherent structure energies of the samples,
 suggesting that the threshold temperature $\TG$ decreases with the inherent
 structure energy (roughly exponentially).   Based on this finding, we
 suggest that experimental ultrastable glasses, that typically lie in
 energy minima below the ones accessible in our numerical simulations,
 would see the anomalies discussed in this work strongly suppressed,
 as in that case $\TG$ would be extremely low or even absent.

\subsection{Aging and heterogeneity of individual samples} 

We now investigate in greater
detail the behavior of individual samples in the regime of times and
temperatures where aging dynamics are slow.  To this end, in addition
to the mean square displacements defined above, we introduce the
displacement of individual particles in two clones,  $u_i(\tw) \propto
|\vec{r}_i^A(\tw) - \vec{r}_i^B(\tw)|^2$, normalized in such a way that
$(1/N) \sum_i \av{u_i(\tw)} = 1$, and following Ref.~\cite{camille2017}
we introduce a susceptibility 
\beq
\chi_{\rm AB}(\tw) = \frac{ \sum_{ij} [ \av{ u_i(\tw) u_j(\tw)} - \av{u_i(\tw)}\av{u_j(\tw)} ] }{ \sum_i [\av{ u_i(\tw)^2} - \av{u_i(\tw)}^2]} \ ,
\eeq
that is equal to 1 if $u_i(\tw)$ and $u_j(\tw)$ are
uncorrelated for all $i \neq j$, while otherwise it gives an estimate
of the correlation length of particle
displacements (raised to an unknown power).
It has been suggested by previous work~\cite{CKPUZ14n,CJPRSZ15,berthier2015growing} that, below the threshold $\TG$,
the system might be ``marginally stable'', i.e. characterized by a diverging correlation length of particle displacements, and a diverging
$\chi_{\rm AB}$, also associated to delocalized soft vibrational modes~\cite{LNSW10,MW15}. However, Ref.~\cite{camille2017} found, in
a system similar to ours, that $\chi_{\rm AB}$ always remains small, suggesting that the low temperature phase is not marginally stable.

\begin{figure}[t]
\centering
\includegraphics[width=\columnwidth]{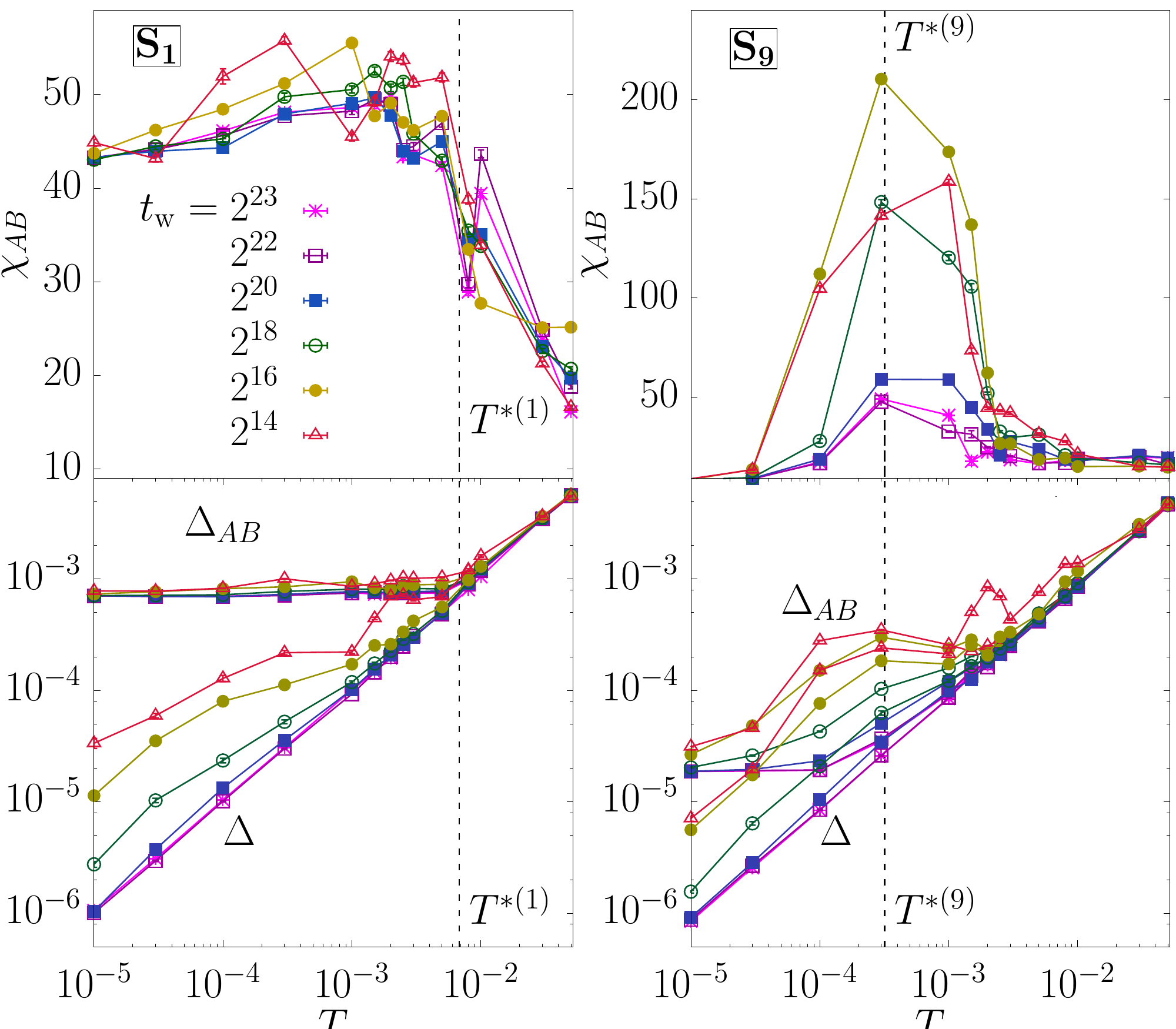}
  \caption{
  {\bf Aging behavior of two representative samples.}
  Values of $\chi_{\rm AB}(\tw)$ (top), and of $\Delta_{\rm AB}(\tw)$, $\Delta(2^{22} + \tw,\tw)$ (bottom) for two representative VD samples obtained with
    $\delta_\mathrm{VD}=2\cdot 10^{-6}$, plotted as a function of temperature $T$ for several values of~$\tw$.
The two samples (labeled 1 and 9) have  $e_{IS}^{(1)}=-3.9522$ and $e_{IS}^{(9)}=-3.9628$. 
Sample 1 displays no aging besides the one related to the convergence of $\Delta(2^{22} + \tw,\tw)$ to its long time limit at low $T$ and only
a moderate increase of $\chi_{\rm AB}$ upon lowering temperature.
Sample 9 displays strong aging around its $T^{*(9)}$, accompanied by a large growth of $\chi_{\rm AB}$.
}\label{fig:susceptibility}
\end{figure}

In Fig.~\ref{fig:susceptibility} we report the aging behavior of $\Delta_{\rm AB}(\tw)$ and $\chi_{\rm AB}(\tw)$ for two individual representative samples,
labeled as sample~1 and sample~9.
In sample~1, we do not observe aging in either $\Delta_{\rm AB}(\tw)$ or $\chi_{\rm AB}(\tw)$, which are independent of $\tw$.  The susceptibility displays only a
moderate increase upon decreasing temperature below $T^{*(1)}$, which is consistent with the results of Ref.~\cite{camille2017}.
In sample~9, instead, we observe strong aging in $\Delta_{\rm AB}(\tw)$ around $T^{*(9)}$, and correspondingly the susceptibility
increases by a factor of about 20 at intermediate times and $T \sim T^{*(9)}$, before relaxing to smaller values at longer times.

\begin{figure}[t]
\centering
\includegraphics[width=\columnwidth]{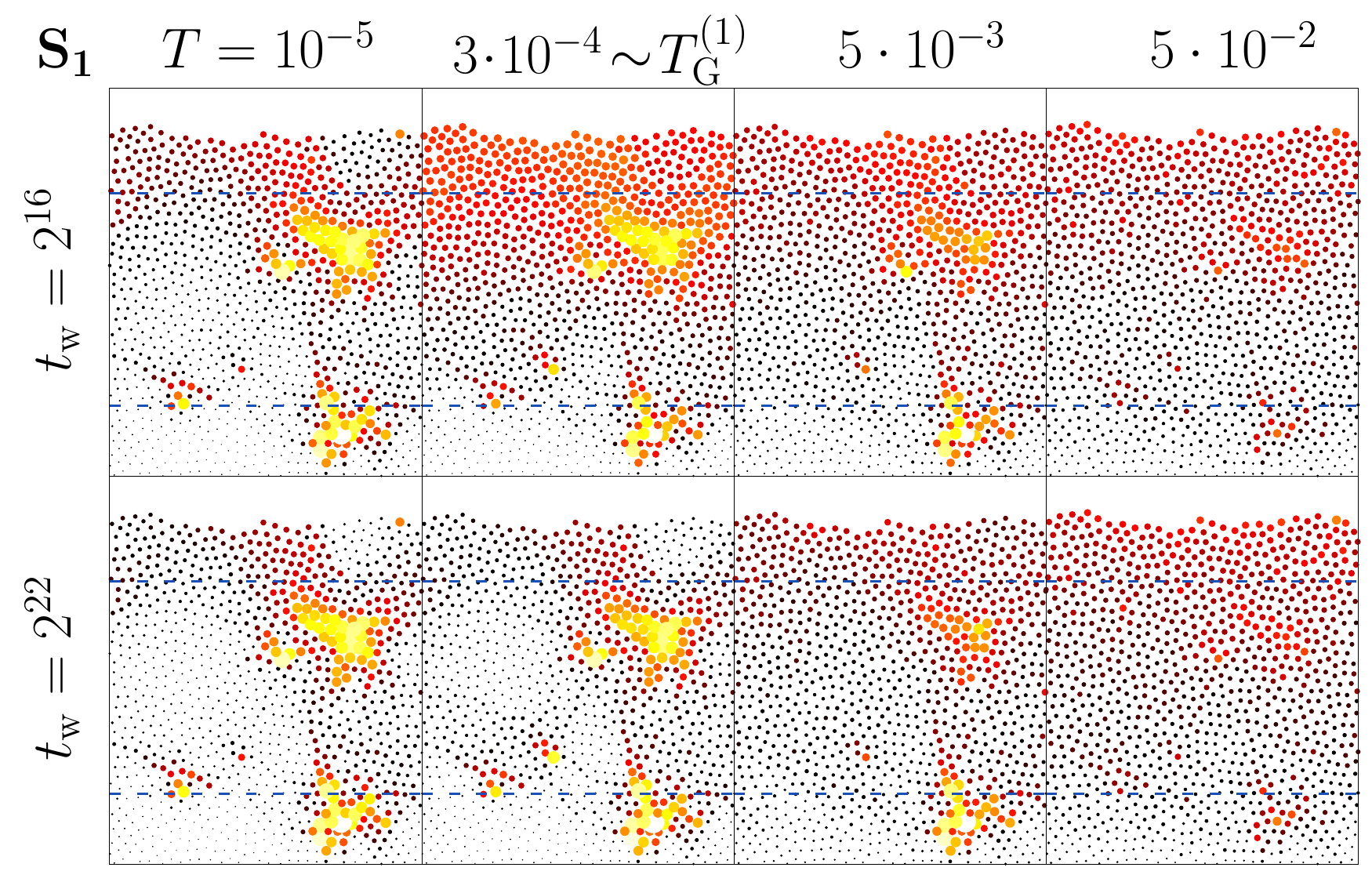}
\includegraphics[width=\columnwidth]{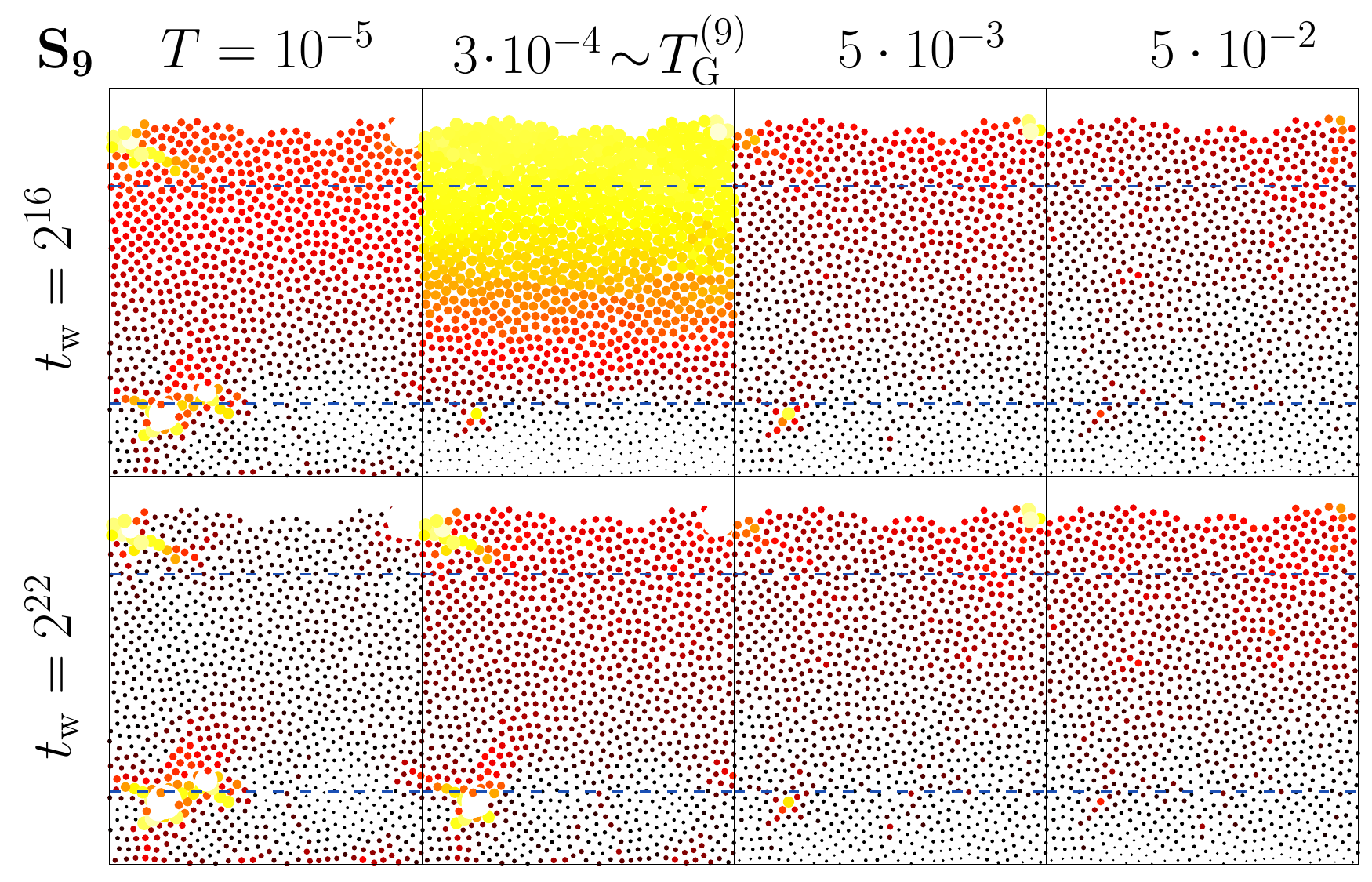}
  \caption{
  {\bf Snapshots of the displacement field.} 
  The displacement field $\av{u_i(\tw)}$, averaged over clones, is shown for the same two representative samples as in Fig.~\ref{fig:susceptibility}.
  For each particle $i$, the corresponding displacement is shown as a circle centered in the particle position, whose area is proportional to
  $\av{u_i(\tw)}$. The colors help visualising the largest displacements. Sample 1 displays only localized defects, while sample 9 displays, at intermediate
  times and $T \sim T^{*(9)} $, a collective displacement of the upper part of the sample.
}\label{fig:snapshots}
\end{figure}

To provide a real space interpretation of these findings, in Fig.~\ref{fig:snapshots} we display snapshots of the displacement
field $\av{u_i(\tw)}$, averaged over clones, for the same two representative samples, at several values of $T$ and $\tw$.
Both samples display, during the aging, a collective displacement of the upper part of the sample, corresponding to a global increase
of density upon cooling --an effect related to the existence of a free surface--, as well as clearly visible localized defects. The main difference between the two samples
is that in sample 9 the surface process leads to greater displacement between clones, indicating that this process happens
in a more heterogeneous way from clone to clone, leading to the stronger aging visible in both $\Delta_{\rm AB}$ and $\chi_{\rm AB}$.
The localized defects are compatible with those observed in Ref.~\cite{camille2017} and lead to a separation of $\Delta^\infty$ and $\Delta^\infty_{\rm AB}$
at low temperatures that is not accompanied by aging nor by a large $\chi_{\rm AB}$. We thus conclude that the system is not marginally stable
below~$\TG$.

\section{Discussion}

We have identified, independently for each sample (or glass basin), a threshold temperature $\TG$, located deep in the glass phase. 
Around this temperature, the aging dynamics after
a quench becomes slow, and vibrational heterogeneity is enhanced. 
Below $\TG$, aging dynamics remains slow, and localized defects
appear, similar to the ones reported in Refs.~\cite{camille2017,HWGM17}. 
The threshold, however, does not correspond to a sharp phase transition and excitations are 
localized below $\TG$. 

Our main result is that $\TG$ markedly decreases with decreasing $E_{\rm IS}$ and thus increasing film stability~\cite{reid2016age,reid2016b}. Hence, ultra-stable glasses with low $E_{\rm IS}$ are also predicted to display a very low $\TG$ and thus remain normal
solids down to extremely low temperatures. 
Our results qualitatively agree with previous studies~\cite{berthier2015growing,camille2017}, but they are obtained
for non-equilibrium films, formed through realistically simulated liquid cooling and physical vapor deposition processes, that
sit higher in the energy landscape~\cite{BCFZ17}.

The theoretical interpretation of our findings is challenging.
Localized defects of 
different nature have been discussed in the context of 
glasses, see e.g.~\cite{LDB16,MSI17,KHGGC11,CWKDBHR10,CSRMRDKL15,JG16,FL98,SWS07,lubchenko2003origin,HKLP10,PRB16},
and our findings could be related to at least some of those theoretical proposals. Future work should clarify these connections, both by additional
numerical simulations and analytical calculations.
The emergence of slow dynamics at low temperature, accompanied by the non-trivial change in the
vibrations of the particles, is reminiscent of the mean field scenario
where these features are consequences of an underlying phase transition, called the
Gardner transition~\cite{Ga85,CKPUZ16}, which separates
a high-temperature
normal solid and a low-temperature marginally stable solid. 
While our results, similarly to the ones of Ref.~\cite{camille2017,HWGM17} suggest that no sharp phase transition takes place in our samples,
one could speculate that the localized excitations we identified are some kind of ``vestige'' of an avoided Gardner-like transition. Because
numerical simulations of hard sphere (colloidal) glasses are instead consistent with the existence of a transition~\cite{berthier2015growing}, it becomes very important
to understand which systems display such a transition and which do not, and why. This is a very important direction for future work, both
analytical~\cite{UB15,LKMY13,AB15,AKLMWY16,CY17}, numerical~\cite{Janus14,CJPRSZ15,berthier2015growing,camille2017,HWGM17} 
and experimental~\cite{seguin2016,GLL17}.

In conclusion,
our observations may explain why some anomalies characteristic of amorphous solids are suppressed in ultra-stable
glasses,
but more work is needed to relate precisely the anomalies observed in our numerically simulated samples to the ones observed in experiments~\cite{QLKMH13,perez2014suppression,LQMKH14,yu2015suppression,TCBSE16}. 
Moreover, finite size effects, and in particular the dependence of our results
on films' thickness, remain to be investigated.

\acknowledgments

  We thank G. Biroli, L. Berthier, M. Ediger, G. Parisi, C. Scalliet,
and P. Urbani for useful exchanges about this work, and
  J. Helfferich for his help at the early step of this work.

This work was granted access to the HPC resources of MesoPSL financed
by the Region Ile de France and the project Equip@Meso (reference
ANR-10-EQPX-29-01) of the programme Investissements d'Avenir
supervised by the Agence Nationale pour la Recherche. Fast
GPU-accelerated codes for simulation of glassy materials were
developed with support from DOE, Basic Energy Sciences, Materials
Research Division, under MICCoM (Midwest Integrated Center for
Computational Materials ).

This project has received funding from the European Union's Horizon
2020 research and innovation program under the Marie
Sk{\l}odowska-Curie grant agreement No. 654971. B.S. was partially
supported through Grant No. FIS2015-65078-C2-1-P, jointly funded by MINECO (Spain) and FEDER (European Union).
This work was supported by a grant from the Simons Foundation
(\#454955, Francesco Zamponi), and by a 
DMREF grant NSF-DMR-1234320.

\appendix

\section{ Details of the system} We work with films of a binary mixture
of $N=1200$, two-dimensional Lennard-Jones particles of type 1 and 2 (where 1 is
more common with concentration $\chi_1\equiv N_1/N\sim 65\%$) that interact with a
third type of particles 3 that act as a fixed substrate at the bottom
of the simulation box. The upper boundary in the vertical axis remains
open and we consider periodic boundary conditions in the direction
parallel to the substrate. The interaction potential between particles
of two species $\alpha,\beta$ separated by a distance $r$ is 
\beq
u_{\alpha,\beta}(r)=4\epsilon_{\alpha,\beta}\caja{\paren{\frac{\sigma_{\alpha,\beta}}{r}}^{12}-\paren{\frac{\sigma_{\alpha,\beta}}{r}}^{6}} \ ,
\eeq
for $r<r_\mathrm{cutoff}^{\alpha,\beta}$, and zero
otherwise.  The cutoff distances are
  $r_\mathrm{cutoff}^{\alpha,\beta}=2.5\sigma_{\alpha,\beta}$, being
  the particle diameters, $\sigma_{11}=1.0 \sigma$,
  $\sigma_{22}=0.88\sigma$, $\sigma_{33}= 0.6\sigma$,
  $\sigma_{12}= 0.8\sigma$, $\sigma_{13}= 0.75\sigma$,
  $\sigma_{23}= 0.75\sigma$. For the potential we use the values
  $\epsilon_{11}=1.0\epsilon$, $\epsilon_{22}=0.5\epsilon$,
  $\epsilon_{33}=0.1\epsilon$, $\epsilon_{12}=1.5\epsilon$,
  $\epsilon_{13}=1.0\epsilon$, $\epsilon_{23}=1.0\epsilon$.  All
  quantities in the paper are shown in Lennard-Jones units, that is:
  $\sigma$, $\epsilon$ and mass $m$ are equal to 1, and time is thus in
  units of $\sigma\sqrt{m/\epsilon}$. Energies in the paper were
measured without shifting the potential to zero at the cutoff
distance, a choice that has no impact during the simulation
considering that updates in the molecular dynamics algorithm only
depend on the derivatives of the interaction potential
$u_{\alpha,\beta}(r)$.  The discrepancies between the inherent
structure energies of this work and the ones shown in
Ref.~\cite{reid2016age} come from the fact that in the previous work
energies were rescaled to compare configurations with exactly the
same portion of type-1 particles in the bulk.  The temperature is
fixed using a Nos\'e-Hoover thermostat~\cite{martyna1992nose} with a temperature damping
parameter $100 \Delta t$, where the time step is here
$\Delta t=0.005$. Inherent structural energies were calculated by
minimizing configurations using the FIRE algorithm with energy and
force tolerances of $10^{-10}$~\cite{bitzek2006structural}.  All simulations were performed
using LAMMPS~\cite{plimpton1995fast}.

Because $\epsilon_{12}$ is higher than $\epsilon_{11}$ and
$\epsilon_{22}$, the most stable configurations tend to maximize the
$1-2$ interactions, which, considering that $1$ particles are more
abundant, tends to displace the particles of type $1$ towards the surface, creating
a clear non-homogeneity along the axis perpendicular to the substrate. In
order to avoid the effect of these two boundaries, all the quantities
computed in this paper were measured using only the particles in bulk,
which corresponds to the central 60\% region (see for
  example the region in between the two horizontal lines in
 Fig.~\ref{fig:snapshots} of the main text). The number of particles in the bulk
varies from sample to sample, but it remains equal to
$N_\mathrm{bulk}\sim 660-670$.


\section{Preparation of glass samples} We prepare glass configurations 
following two
distinct protocols: slow cooling from the liquid phase (SC) with two distinct cooling rates $\delta_{\mathrm{SC}}=2\cdot 10^{-4}$ and
$\delta_{\mathrm{SC}}=2\cdot 10^{-9}$ down to $T_{\rm f}=0.05$,
and a protocol mimicking the vapor
deposition (VD) procedure using four different particle-deposition
rates $\delta_{\mathrm{VD}}$ with substrates at different temperatures $T_{\rm s}$.  
The details concerning this protocol 
can be found in Ref.~\cite{reid2016age}; we selected for each deposition
rate $\delta_{\mathrm{VD}}$ the value of $T_{\rm s}$ that corresponds to the lowest inherent structure energy of the
resulting glass,
which gives $T_{\rm s}=0.18$ at $\delta_{\mathrm{VD}}=2\cdot 10^{-3}$,
$T_{\rm s}=0.16$ at $\delta_{\mathrm{VD}}=2\cdot 10^{-5}$, $T_{\rm s}=0.14$ at $\delta_{\mathrm{VD}}=2\cdot 10^{-6}$, 
and $T_{\rm s}=0.16$ at $\delta_{\mathrm{VD}}=2\cdot 10^{-7}$.

Following each
protocol, we prepare $N_\mathrm{s}$ independent glasses (to which we
will refer here as {\em samples}), each corresponding to a distinct glass basin in the energy landscape.  
We have considered
$N_\mathrm{s}=10$ for all the cases with the exception of the VD
glasses obtained with the slowest particle deposition rate, where only
$5$ samples were considered. 
We define the inherent structure (IS) of a configuration
as the energy minimum that is reached by minimizing the energy starting from that
configuration~\cite{SW82}.
The different protocols allow us to
produce glasses with a wide range of inherent structure energies
$E_\mathrm{IS}$ (Fig. 1b in the main text).

\section{Cloning procedure} 
To study the vibrational anomalies of a glass basin,
for each of these samples, we
create $N_\mathrm{c}$ {\em clones}, which correspond to different
configurations of the same glass state, using the following procedure:
\begin{itemize}
\item We first cool the initial configuration instantaneously to $T=T_{\rm clone} =0.05$, 
  and let it relax until we observe no more aging in the height of the plateau (during $2^{14}$  
  time steps). 
This temperature is chosen because for $T<T_{\rm clone}$,
$E_\mathrm{IS}$ becomes independent of temperature for all
the samples, and furthermore no diffusion is observed at $T_{\rm clone}$ (with the exception of the samples prepared by the fastest cooling and deposition rates, i.e. $\delta_\mathrm{SC}=2\cdot 10^{-4}$ and  $\delta_\mathrm{VD}=2\cdot 10^{-3}$ respectively, where some diffusion is still observed at this temperature at long times).
These two observations imply that at $T_{\rm clone}$ the configurations are trapped into well-defined
glass basins, which is not always the case at the preparation temperature ($T_{\rm f}$ or $T_{\rm s}$ depending on the protocol), where residual
diffusion and inherent structure energy variations are observed in some samples.
\item
Stable glass configurations obtained at $T_{\rm clone}$ are
then cloned by performing $N_\mathrm{c}=200$ short
independent simulations assigning to each configuration a set of independent random velocities
drawn from the Maxwell distribution
at $T_{\rm clone}$, as shown in the inset of
Fig.~1 in the main text.  The length of these simulations is chosen
to be longer than the ballistic regime, to let the particles explore
their inner cages (of average sizes $\Delta^\infty$). 
In our case, $2^{10}$ 
(the vertical dotted line in Fig.~\ref{fig:eIreid}a in the main text) satisfied these requirements for
all our samples.
\end{itemize}
The clones thus represent independent configurations of a same sample at the cloning temperature $T_{\rm clone}$.

\section{Instantaneous quenches in temperature} Now starting from each
of these clones, we perform instantaneous quenches to lower
temperatures. That is, we rescale the velocities of the particles
in such a way that the kinetic energy corresponds to a temperature $T < T_{\rm clone}$,
and then 
we use standard
molecular dynamics to follow the evolution of the system,
keeping the temperature fixed by
a Nos\'e-Hoover thermostat~\cite{martyna1992nose}. The initial time corresponds to the time of the quench, and we call $\tw$
the time elapsed since the quench.

\bibliography{biblio,HS}

\end{document}